# Metadata in SDN API

Dmitry Namiot

Faculty of Computational Mathematics and Cybernetics
Lomonosov Moscow State University
Moscow, Russia
dnamiot@gmail.com

Manfred Sneps-Sneppe

Ventspils International Radioastronomy Centre
Ventspils University College
Ventspils, Latvia
manfreds.sneps@gmail.com

*Abstract*—**This paper discusses the system aspects of development of applied programming interfaces in Software-Defined Networking (SDN). Almost all existing SDN interfaces use so-called Representational State Transfer (REST) services as a basic model. This model is simple and straightforward for developers, but often does not support the information (metadata) necessary for programming automation. In this article we cover the issues of representation of metadata in the SDN API.**

*Keywords—SDN; northbound API; REST; metadata; WSDL; Parlay;*

## I. INTRODUCTION

Software-Defined Networking (SDN) is an emerging architecture for networks. SDN offers dynamic, manageable, cost-effective, and adaptable solutions for networks architecture [1]. This approach separates the network control and forwarding functions. SDN enables the network control to be directly programmable (to be software defined). It offers also the abstract model for the network infrastructure for applications and network services [2].

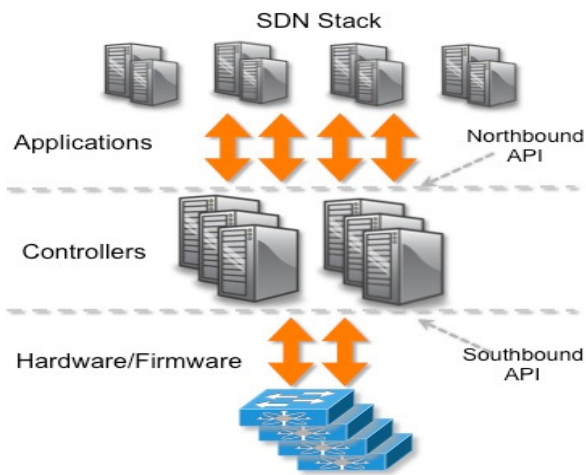

Fig. 1. SDN Stack [3]

This switches promise to break physical boundaries on network equipment through precisely defined Application Programming Interfaces (APIs). So, in SDN the whole networking process is software driven. API is the key point of this new paradigm. Figure 1 illustrates the various APIs in SDN.

As soon as APIs form the main engine behind SDN, API management becomes a critical component to build, manage and scale applications for our networks. One important aspect of this management is automation for programming. The ability to automate coding lets save time, at the first hand [4]. And time to market is the most important feature for any new development. Such automation lets also reuse existing solutions and leads to the more robust and bugs-free systems.

The ability to automate the work process for any API directly depends on the ability to obtain programmatically all information about this API. In other words, API should support metadata. In this paper we would like to talk about several aspects on so-called Northbound API. As the most of modern APIs, it is some REST [5] based solution (actually, solutions). And metadata support in REST services has its own characteristics.

The rest of the paper is organized as follows. In Section II we discuss Northbound API. In Section III we discuss REST approach and metadata in REST. And Section IV is devoted to discussion.

## II. NORTHBOUND API IN SDN

So-called Northbound API forms the application layer for SDN controller and provides the main support for 3-rd party applications (Figure 2):

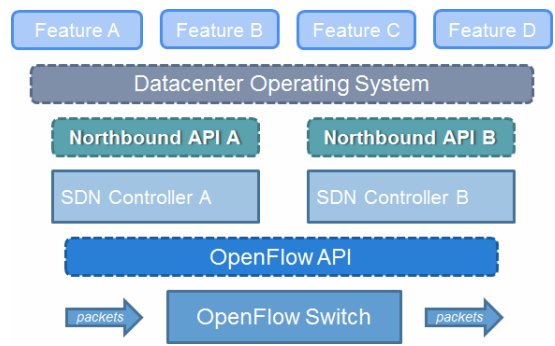

Fig. 2 Northbound API [6]

Actually, the term Northbound API may cover the whole set of APIs. It is for the different layers of abstraction for SDN controller. One application (e.g., email) should not require much detailed information about the SDN (the underlying network). So, Northbound API can present for such applications a high degree of abstraction. Other applications (e.g., firewall) would require far more detailed (elaborated) network information from the SDN. It will be served via own level of abstraction in Northbound API. There is Northbound API working group [7] whose task is to develop a standard (standards).

At this moment (The Working Group has not yet completed the design), we can be sure that Northbound API is REST based [8]. REST as an architectural style is the most used approach for the modern services [9]. As the main features which attract developers, we can name here the uniform interface, statelessness and self-descriptive messages.

But let us see one example from Neutron API (OpenStack) [10]. The request is a typical REST:

```
GET /v2.0/networks?limit=2

Accept: application/json
```

And here is a response (part of it):

```
{

"networks":[

{

"status":"ACTIVE",

"subnets":[

"a318fcb4-9ff0-4485-b78c-9e6738c21b26"

],

"name":"private",

"admin_state_up":true,

"tenant_id":"625887121e364204873d362b55
3ab171",

"id":"9d83c053-b0a4-4682-ae80-
c00df269ce0a",

"shared":false

}

]

}
```

Uniform interface in REST means that all resources present the same interface to clients. For comparison, the SOA approach [11] may offer personalized interfaces for the different resources. Statelessness means that the server does not save state on the client's behalf. By this reason, all requests in REST must carry the session-oriented information [12]. This principle enables caching for REST services. Resource representation (naming) is also a basic principle of REST. Resources are named with unique identifiers. Clients use these identifiers for interaction. All this is very convenient to the developers.

In the same time, the Service Oriented Architecture (SOA) model which is often compared with REST is based on the idea that different services have different interfaces. It means, immediately, that we need to provide the definition for used interfaces. Indeed, the definition for the services is a key part of SOA. For example, Web Service Definition Language (WSDL) [13] is a part of SOA specification. A WSDL definition of a Web Service defines operations in terms of their underlying input and output messages. Unlike this, REST is based on the self-described messages. Indeed, WSDL defines the form of the data that accompany the messages in SOA. REST does not provide this information. In other words, SOA has got a rich set of metadata.

The problem with meta-data support is very transparent. Let us return to the above mentioned Neutron API example. How the developers can get information ("get" means programmatically discovery) about the following elements:

HTTP command (it is GET in this case)

URI (it is /v2.0/networks)

Output formats (it is JSON)

Version (it is 2.0 in this case)

Optional and mandatory parameters (it is limit=2)

There is no ways to discover this information programmatically in the modern implementations of REST approach. And without the metadata there is no way to automate programming [14]. That is why we can conclude that metadata for Northbound API is a key problem.

III. METADATA FOR REST APPROACH

Let us see what kind of metadata could be useful for REST models.

*WSDL.* We can use WSDL as a base for our research. We can suggest the following list of elements:

1. Endpoints specifications. Here we should be able to describe URLs for requests, as well as resource paths.

2. Methods for access to these resources. Actually, it is a list of supported HTTP commands (GET, POST, etc.).

3. Queries (parameters) for requests. We should be able to specify also the mandatory and optional parameters.

4. HTTP headers

5. Status codes and error messages

6. Versioning support.

7. The formats for the responses. For example, it could be JSON or XML.

For some items from this list, we can mention the solutions (agreements) in general use. For example, version info could be presented as a part of URI. For the above mentioned example (Neutron API) it is */v2.0.* The format for responses could be also presented as a part of URI. E.g., */networks.json* for JSON output and */networks.xml* for XML.

The deeper level of software description could be achieved via Web Ontology Language (OWL) [15]. In general, the knowledge about structural and behavioral properties of software can be shared across the software engineering community in the form of design patterns expressed in the OWL [16]. In this connection, we can mention several papers devoted to OWL-based description of public Web APIs (Twitter API [17], Google+ API [18], etc.). The goal is to simplify the development for mashups [19]. This use case could be a perfect fit for Northbound SDN API. Software API for network equipment opens the new direction for telecommunications mashups.

Technically, we can use WSDL 2.0 for metadata in REST models [20]. The root element of a WSDL 2.0 document is the *description* element. It has the four child elements:

*types*

*interface*

*binding*

*service*

The *types* element contains all the XML schema elements and type definitions that describe the service's messages. As per WSDL 2.0 spec, we can use other type systems (e.g., JSON).

The *interface* element defines the Web service operations. It includes the specific input, output, and fault messages. The *interface* element defines the order in which messages are passed.

The *binding* element defines how a client can communicate with the Web service. For REST services we can specify HTTP here.

The *service* element associates an address for the Web service with a specific *interface* and *binding*. There are also two additional namespaces useful REST (HTTP and extension).

But overall, this approach has not been adopted by the development community. According to the dominant ideology, REST messages must be self-describing.

***API Blueprint***. Actually, this idea with messages describing prevails in the existing products for metadata support in REST models. We can list here several products with the common conceptual idea – take the annotated code and create automatically (semi-automatically) manuals (test suite) for REST services. For example, it is API Blueprint [21]. API Blueprint is a documentation-oriented API description language. The API Blueprint is essentially a set of semantic assumptions laid on top of the Markdown syntax used to describe a Web API. A blueprint is a plain text Markdown document describing a Web API or its part. Here is an example of the API Blueprint (Apiary.io):

```
## Folder [/folder{id}]

A single Folder object, it represents a
single folder.
```

```
Required attributes:

- `id` Automatically assigned

- `name`

- `description`

Optional attributes:

- `parent` ID of folder that is the
parent. Set to 0 if no parent

+ Parameters

+ id (required, int) ... Unique folder
ID in the form of an integer

  + Model (application/hal+json)
```

The automatically generated documentation looks like a readable (and ready for web publishing) text with descriptions like this:

```
Function: Retrieve a single Folder

Command: GET

Path: /folder{id}

Parameters Name:      id

Description   Details:  Unique  folder
ID in the form of an integer

Type:      int, required.
```

The similar model is supported by the I/O Docs system [22].

***RAML.*** As the next step, we should mention several applications for so-called API management. The typical example is API backing platform RAML [23]. The idea is to simplify the entire API lifecycle. This system covers the development from design to management and publishing. It has API Designer as a web based tool to help developers quickly design RESTful APIs. Once the API is designed, developers can use the API Notebook to explore the API, test it, and iteratively improve it [24]. Using code-generator, developers can build API. There is a simple framework that automatically generates an integrated structure for proposed API, so developers can hook into your services and data sources. The API could be immediately deployed to the API Gateway, a dedicated orchestration layer. Once a new API is running, it is possible to use API Manager as a single point of management for all APIs to control access, enforce SLAs, and monitor users and traffic to keep APIs and services running at peak performance.

The biggest problem with such systems is the lack of openness. Actually, they should have a special layer (adaptors) for connected APIs. In particularly, developers will need an adaptor for Northbound SDN API too.

***WADL.*** The most promising attempt (by our opinion) to add metadata support to REST services is WADL [25]. It is an XML-based file format that provides a machine-readable description of HTTP-based Web applications. It was submitted to the W3C but has not been standardized on (as far as we know). It is the direct "equivalent" of Web Services Description Language. WADL is intended for applications that are based on the Web architecture. Like WSDL, it is a language independent platform. It aims to promote reuse of applications beyond the basic use, inside a Web browser. WADL models the resources provided by an embedded device and the relationships among them. Here is an example:

```
<request>
    <headers>
        <header required="true|false">
            <name />
            <description />
            <value />
        </header>
    </headers>
    <url>
        <parameters_set>
            <name />
                <value />
            </parameters_set>
    </url>
</request>
```

There are several other projects like RSDL (RESTful Service Description Language) [26] or JSON Hyper Schema (a vocabulary for JSON Schema to describe links) [27], but none of them has been adopted by developers.

## IV. DISCUSSION

Our analysis and our practical experience show that at this moment the above-mentioned instruments for metadata management in REST models are not used. Almost always, REST model operates without metadata, except the manual for developers only.

Some of the above-mentioned markdown-based solutions are used in the best case to generate documents (manuals). The use of integrated systems (e.g., the above-mentioned RAML) is associated with a particular manufacturer and depends on the availability adaptors for APIs. Independent tools (e.g., WADL), for unknown reasons, have not received the approval of the developers.

However, we think the formal description (the formal presentation) for any public API (including Northbound SDN API) is very important. Of course, the idea of a public API for network nodes is not entirely new. As a direct analogue, we can mention Parlay, for example. Telephone switching system (PBX) is also a network node, and Parlay (Parlay X) API [28] plays the same role as the Northbound SDN API. It is a public

API, provides an open programming interface to network hardware for application programs (for applied services). And lessons from Parlay's failure are also important. Its main problems were connected with the difficulty of adopting API to the developers [29]. The development of services with Parlay API, in fact, did not save development time. Vice versa, it takes usually more time than the deployment of non-portable (closed) APIs. Time to market is a key indicator for software development tools. The first (and possibly the main) purpose of using metadata of some API is automation for software development. So, it is directly affects the key indicator (time to market).

In connection with this, the next question seems very correct. Shall we add some common metadata support solution to the upcoming specification of Northbound SDN API? As seems to us, it would be possible to choose a suitable form of metadata support (e.g., WADL). It should be suitable for developers of SDN controllers. And then, each version (implementation) of Northbound API would be accompanied with its own version of metadata descriptions. This is a reproach to the process of adaptation Northbound SDN API, and it would serve the interests of all developers of applied services. Application developers are the real users for Northbound SDN API.